\documentclass[12pt]{IEEEtran}
\usepackage{a4wide}
\usepackage{graphicx}
\usepackage{url}

\begin{document}

\title{Shawn: A new approach to simulating wireless sensor networks}

\author{
  \authorblockN{
  A.~Kr{\"o}ller\authorrefmark{1},
  D.~Pfisterer\authorrefmark{2},
  C.~Buschmann\authorrefmark{2}},
  S.~P.~Fekete\authorrefmark{1},
  S.~Fischer\authorrefmark{2}
  \\[1ex]
  \authorblockA{\authorrefmark{1}
    Institute of Mathematical Optimization,\\
    Braunschweig University of Technology,\\
    D-38106 Braunschweig, Germany,\\
    E-Mail: \{a.kroeller,s.fekete\}@tu-bs.de\\[1ex]
  }
  \authorblockA{\authorrefmark{2}
    Institute for Telematics,\\
    University of L{\"u}beck,\\
    D-23538 L{\"u}beck, Germany,\\
    E-Mail: \{pfisterer,buschmann,fischer\}@itm.uni-luebeck.de
  }
}

\date{January 31, 2005}

\maketitle

\begin{abstract}
  We consider the simulation of wireless sensor networks (WSN) using a
  new approach. We present Shawn, an open-source discrete event 
  simulator that has considerable differences to all other
  existing simulators. Shawn is very powerful
  in simulating large scale networks with an abstract point of view.
  It is, to the best of our knowledge, the first simulator to 
  support generic high-level algorithms as well as distributed protocols
  on exactly the same underlying networks.
\end{abstract}

\section{Introduction}
\label{sec:intro}
In recent times, the study of wireless sensor networks (WSN) has become
a rapidly developing research area that offers fascinating
perspectives for combining technical progress with new applications of
distributed computing. Typical scenarios involve a large swarm of
small and inexpensive sensor nodes, each providing limited computing
and wireless communication capabilities that are distributed in some
geometric region. From an algorithmic point of view, the
characteristics of sensor networks require the shift to a new paradigm
that is different from classical models of computation: The absence of
centralized control, limited capabilities of nodes and low bandwidth
communication between nodes require developing new algorithmic ideas
that combine methods of distributed computing and network protocols
with traditional centralized network algorithms.

To acquire a deeper understanding of these networks, three
fundamentally different approaches exist: Analytical methods, computer
simulation, and physical experiments.  Designing algorithms for sensor
networks can be inherently complex. Many aspects such as energy
efficiency, limited resources, decentralized collaboration, fault
tolerance, and the study of the global behavior emerging from local
interactions have to be tackled.

In principle, experimenting with actual sensor networks would be a
good way of demonstrating that a system is able to achieve certain
objectives, even under real-world conditions. However, this approach
poses a number of practical difficulties. First of all, it is
difficult to operate and debug such systems. This may have
contributed to the fact that only very few of these networks have yet
been deployed \cite{mainwaring-habitat,analysisgdi2,zebranet}.
Real-world systems typically consist of roughly a few dozen sensor
nodes, whereas future scenarios anticipate networks of several
thousands to millions of nodes~\cite{estrinembed,atomiccomputing}.
Using intricate tools for simulating a multitude of parameters, it may
be possible to increase the real-world numbers by a couple of orders
of magnitude. However, the difficulty of pursuing this approach
obfuscates and misses another, much more crucial issue: Designing
highly complicated simulation tools for individual sensor nodes
resembles constructing a working model for individual brain cells.
However, like a brain is much more than just a cluster of cells,
realizing the vision of an efficient, decentralized and
self-organizing network cannot be achieved by simply putting together
a large enough number of sensor nodes. Instead, coming up with the
right functional structure is the grand scientific challenge for
realizing the vision of sensor networks. Understanding and designing
these structures poses a great number of algorithmic tasks, one level
above the technical details of individual nodes. As this understanding
progresses, new requirements may emerge for the capabilities of
individual nodes; moreover, it is to be expected that the technical
process and progress of miniaturization may impose new parameters and
properties for a micro- simulation.

\subsection{Motivation of this work}
\label{sec:motivation}
Consider the situation of developing localization algorithms. 
Usually one is interested in the quality of the solution that is
produced by a specific algorithm. There is certainly some influence of
communication characteristics, e.g., because they may affect
transmission times and hence communication paths and loss. From the
algorithm's point of view, there is no difference between a complete
simulation of the physical environment (or lower-level networking
protocols) and the alternative approach of simply using well-chosen
random distributions on message delay and loss. This means that
using a detailed simulation may lead to the strange situation in which
the simulator spends much processing time on producing results that are of no
interest at all, thereby actually hindering productive research on the
algorithm.

This is the central idea of the proposed simulation framework: By
replacing low-level effects with abstract and exchangeable models, the
simulation can be used for huge networks in reasonable time.
Section~\ref{sec:casestudy} shows the speedup that we achieve by
replacing the popular simulator Ns-2~\cite{ns-2} with our own
simulator Shawn~\cite{shawn}.

Shawn is licensed under the GNU General Public License. It is
available for download at \url{http://www.swarmnet.de/shawn}.

The rest of the paper is organized as follows:
Section~\ref{sec:related} categorizes available simulation tools that
cover the simulation of sensor networks. Section~\ref{sec:goals}
discusses the differences to the Shawn simulator presented in this
paper. The overall architecture of Shawn is presented in
Section~\ref{sec:arch}. Section~\ref{sec:casestudy} serves as an
example on how users can benefit from Shawn. In
Section~\ref{sec:conclusion} we summarize the scientific contributions
of our approach and the conclusions that can be drawn. In
Section~\ref{sec:future} we discuss our plans for further development
of the Shawn project.
\section{Related Work}
\label{sec:related}
The range of applications for simulation is rather broad and many
simulators have been developed in the past. Each of them targets a
specific application domain in which it can deliver best results. The
semantics of what is actually meant by the term ``simulation'' varies
heavily among researchers and publications, depending on the goals of
the simulations in question.

\begin{figure*}[t]
  \centering
  \includegraphics[width=.6\textwidth]{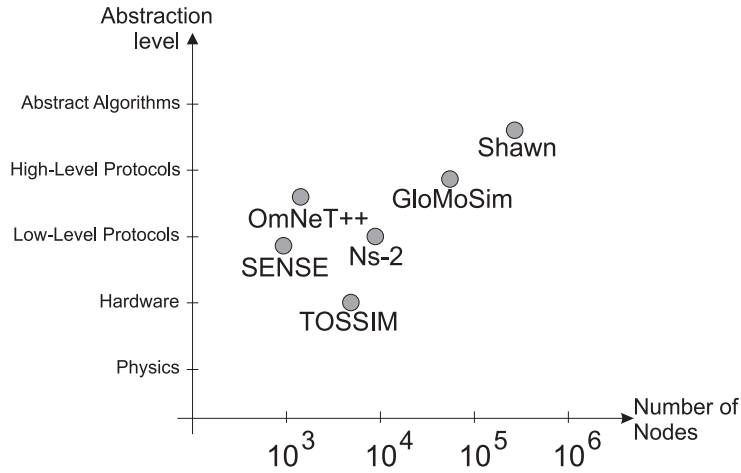}
  \caption{Intended application area of simulators.}
  \label{fig:class}
\end{figure*}

This often results in the simulation of physical phenomena such as
radio signal propagation characteristics and ISO/OSI layer protocols,
e.g., media access control (MAC). Other approaches focus on algorithmic
aspects and they abstract from lower layers. The first approach delivers a
precise image of what happens in real networks and how the protocols
interact with each other at the cost of resource-demanding simulations,
leading to scalability problems. The latter type employs abstract
models of the real world, instead of simulating it down to the bit
level. Important questions are the analysis of the network structure
as well as the design and evaluation of algorithms (and not
protocols). We have coarsely categorized some of the most prominent
simulation frameworks according to the criteria of scalability and
abstraction level.  Figure~\ref{fig:class} classifies the application area of these
simulators along two axes, showing abstraction level and number of
network nodes. Note that this does not express the maximal feasible
network sizes, but rather reflects the typical application domain. For
example, nearly every simulator can handle huge networks if the
connectivity is kept near zero, which does not help in choosing the
appropriate simulator for a given task.

We now give an overview of different simulators that are commonly used
for sensor networks.

\paragraph{Ns-2~\cite{ns-2}}
The ``Network Simulator-2'' is a discrete event simulator targeted at
network research. It is probably the most prominent network simulator.
It includes a huge number of protocols, traffic generators and tools
to simulate TCP, routing, and multicast protocols over wired and
wireless (local and satellite) networks. Its main focus is the ISO/OSI
model simulation, including phenomena on the physical layer and energy
consumption models. Ns-2 features detailed simulation tracing and
comes with the simulation tool ``network animator'' (nam) for later
playback. It is available for free under an open source license.
Support for sensor network simulations has also been integrated
recently \cite{ns2sensors,sensorsim}, including sensing channels,
sensor models, battery models, lightweight protocol stacks for
wireless micro sensors, hybrid simulation support and scenario
generation tools.  The highly detailed packet level simulations lead
to a runtime behavior closely coupled with the number of packets that are
exchanged, making it virtually impossible to simulate really large
networks. In principle, Ns-2 is capable of handling up to 
16,000 nodes, but the level of detail of its simulations 
leads to a runtime that makes it hopeless to deal with more than 1,000
nodes. Ns-2's long
development path since 1989 has led to a vast repository for network
simulations but also reflects its downside: It has a steep learning
curve and requires advanced skills to perform a meaningful and
repeatable simulation. The diverse distributions of Ns-2 that are used by
research groups around the world complicate the comparability of
achieved results.

\paragraph{OMNeT++ \cite{omnet++}}
The ``Objective Modular Network Testbed in C++'' is an object-oriented
modular discrete event simulator. Like Ns-2, it also targets the
ISO/OSI model. It can handle thousands of nodes and features a
graphical network editor and a visualizer for the network and the data
flow. The simulator is written in C++ for high performance and comes
with a homegrown configuration language ``NED''. OMNeT's main
objective is to provide a component architecture through which
simulations can be composed very flexibly.  Components are programmed
in C++ and then assembled into larger components using NED. It is free
for academic purposes, a commercial license is also available.

\paragraph{GloMoSim \cite{GloMoSim}}
The ``Global Mobile Information Systems Simulation Library'' is a
scalable simulation environment for wireless and wired network
systems. It is modeled on the ISO/OSI principle using a layered
design.  Standard APIs are used between the different simulation
layers (Application, Transport, Network, Data Link, Packet Reception
Models, Radio Model, Radio Propagation and Mobility). Though
potentially designed for simulations with 100,000 nodes, just 5,000
nodes already lead to runtimes of about an hour on a single machine.
However, support for parallel execution is provided. The simulator is
built on top of Parsec~\cite{Parsec}, which provides the parallel
discrete event simulation capabilities.  Though also designed for
wired networks, GloMoSim currently supports only protocols for the
simulation of purely wireless networks.

\paragraph{SENSE \cite{SENSE}}
This is a simulator specifically developed for the simulation of
sensor networks.  It offers different battery models, simple network
and application layers and a IEEE 802.11 implementation. With regard
to scalability, the authors plan to enable SENSE to allow for
parallelization in the future. In its current version, SENSE comes
with a sequential simulation engine that can cope with around 5,000
nodes, but depending on the communication pattern of the network this
number may drop to 500. The authors identify extensibility,
reusability and scalability as the key factors they address with
SENSE. Extensibility is tackled by avoiding a tight coupling of
objects by introducing a component-port model, which removes the
interdependency of objects that is often found in object-oriented
architectures. This is achieved by their proposed ``simulation
component classifications''. These are essentially interfaces, which
allows exchanging implementations without the need of changing the
actual code. Reusability on the code level is a direct consequence of
the component-port model.

\paragraph{TOSSIM \cite{TOSSIM}}
The ``TinyOS mote simulator'' simulates TinyOS~\cite{TinyOS} motes at
the bit level and is hence a platform-specific simulator/emulator. It
directly compiles code written for TinyOS to an executable file that
can be run on standard PC equipment. Using this technique, developers
can test their implementation without having to deploy it on real
sensor network hardware. TOSSIM can run simulations with a few
thousand virtual TinyOS nodes.  It ships with a GUI (``TinyViz'')
that can visualize and interact with running simulations. Just
recently, PowerTOSSIM~\cite{PowerTOSSIM}, a power modeling extension,
has been integrated into TOSSIM. PowerTOSSIM models the power consumed
by TinyOS applications and includes a detailed model of the power
consumption of the Mica2~\cite{Mica2Mote} motes.

\paragraph{BOIDS}
In the context of the BOIDS project~\cite{boidsproject}, a number of
simulation and visualization tools have been developed. BOIDS reaches
back to 1987 and studies the global behavior of a group of mobile
individuals emerging from their local interaction. The authors model
reciprocity as so-called steering behaviors \cite{reynolds99steering},
an abstract concept similar to attractive and repelling forces.
However, these tools must be considered visualizers for bio-inspired
agent behavior, rather than full-scale network simulators.

The crucial point of the above listing is that each of the
simulators has its area of expertise in which it excels.
Unfortunately, none of these areas happens to be high-level protocols
and abstract algorithms in combination with the speed to handle large
networks. This is the gap that is filled by Shawn.

\section{Shawn Design Goals}
\label{sec:goals}
Shawn differs in various ways from the other simulators. The most
notable difference is the focus of interest that is covered. Shawn does
not try to compete with the other simulators in the area of network
stack simulation: As already described, we do not believe that this is 
a fruitful approach for the evaluation of protocols and algorithms for 
wireless sensor networks. The behavior of the network as
a whole should be modeled in a way that allows for the needed
performance and development speed. Our main focus is to support the
steps that are necessary in order to achieve a complete protocol
implementation. For this purpose, various algorithmic
preliminary considerations are necessary.

The following subsections discuss several aspects in which Shawn
differs significantly from other existing simulation frameworks by
pointing out the main design paradigms of Shawn.

\subsection{Simulating the effects}
\label{sec:simeffects}
One central approach of Shawn is to simulate the effect caused by a
phenomenon, not the phenomenon itself. For example, instead of
simulating a complete MAC layer including the radio propagation model,
its effects (i.e., packet loss and corruption) are modeled in Shawn.
This has several implications for simulations: They get more
predictable and meaningful, and there is a huge performance gain,
because such a model can often be implemented very efficiently. This
also results in the inability to come up with the detail level that,
say, Ns-2 provides with respect to physical layer or packet level
phenomena.

We are convinced that modeling network characteristics,
such as increased packet loss triggered by high traffic, yields
equivalent results compared to calculating possible congestion for
single packets, while offering a number of advantages.  
For example, when using a simplified communication model in
simulations of a localization algorithm, the quality of solutions is
only slightly affected. On the other hand, running times are not
comparable at all.

This distinction is the underlying paradigm of our large-scale
high-speed simulation environment: It makes sense to simplify the
structure of some low-level parameters: Their time-consuming
computation can be replaced by fast simulation, as long as the
interest in the large-scale behavior of the macro-system focuses on
unaffected properties.

\subsection{Simulation of huge networks}
\label{sec:hugenets}
One direct benefit of the above paradigm is superior
scalability. Visionary scenarios anticipate networks with a huge
number of individual nodes. It is to be expected that these networks will
consist of potentially millions of nodes
\cite{estrinembed,atomiccomputing}, so a simulator must be capable
of operating with that many nodes. One critical issue in
designing Shawn was to support node numbers orders of magnitudes higher 
than the currently existing simulators. We have successfully
run simulations on standard PC equipment with more than 100,000 nodes.

\subsection{Supporting a development cycle}
\label{sec:/dev/cycle}
Shawn inherently supports the development process with a complete development
cycle, beginning at the initial idea, ultimately leading to a fully
distributed protocol. In the following the complete development cycle
of simulations using Shawn is depicted, with 
each step being optional.

Given a first idea for an algorithm, it is natural to assume that the
next step is not to design some protocol, but to perform a structural
analysis of the problem at hand. To get a better understanding of the
problem in this first phase, it may be helpful to look at some example
networks and analyze the network structure and underlying graph
representation.

In order to achieve a rapid prototype version, the next step is to
implement a first centralized version of the algorithm. A centralized
algorithm has full access to all nodes and has a global, flat view of
the network. This provides a simple means to get results and a first
impression of the overall performance of the examined algorithm.  The
results emerging from this process can provide optimization feedback
for the algorithm design.

Once a satisfactory state of the centralized version has been
achieved, the feasibility of its distributed implementation can be
investigated in depth. Only a simplified communication model between
individual sensor nodes is utilized at this point in time. Because the
goal of this step is to prove that the algorithm can be transformed to
a distributed implementation, the messages exchanged between the nodes
are simple data structures passed in memory. This allows for a very
efficient and fast implementation, leading to meaningful results.

Having arrived at a fully distributed and working implementation, the
remaining task is to define the actual protocol and rules for the
nodes to run the distributed algorithm. Messages that have been
in-memory data structures that are passed as references may now be
represented in form of individual data packets. With the protocol and
data structures in place, the performance of the distributed
implementation can be evaluated. Interesting questions that can be
explored are, e.g., the number of messages, energy consumption,
run-time, resilience to message loss and environmental effects.

\section{Architecture}
\label{sec:arch}
\begin{figure*}[t]
  \centering
  \includegraphics[width=.8\textwidth]{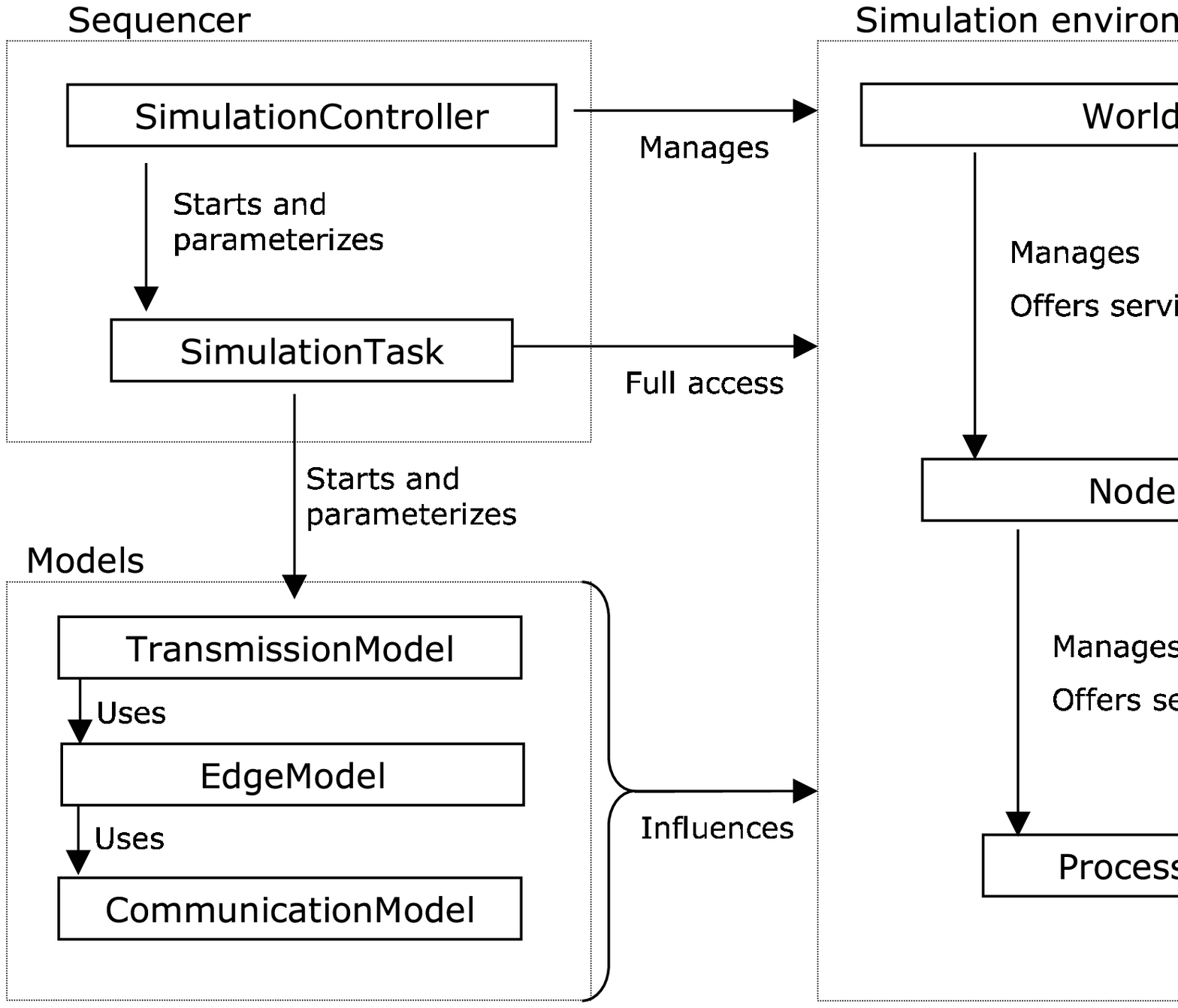}
  \caption{Architectural overview of Shawn's core components.}
  \label{fig:arch}
\end{figure*}

Conceptually, Shawn consists of three major parts: Simulation
environment, Sequencer, and Models. The simulation environment contains
the simulated items and their properties, while the sequencer and the
models influence the behavior of the simulation environment.
Figure~\ref{fig:arch} shows a high-level overview of the architecture
of Shawn.

\subsection{Models}
\label{sec:models}

To achieve reusability, extensibility and flexibility, exchangeable
models are used wherever possible in Shawn. A thorough distinction
between models and their respective implementations supports these
goals. Shawn maintains a very flexible and powerful repository of
model implementations that can be used to compose simulation setups
simply by selecting the desired behaviors through model identifiers at
runtime.

Some models shape the behavior of the virtual world, while others
provide more specialized data. Models that form the foundation
of Shawn are the {\em Communication Model}, the {\em Edge Model} and
the {\em Transmission Model}.

The {\em Communication Model} determines for a pair of nodes whether
they can communicate.  There may be models representing unit disk
graphs for graph-theoretical studies, models based on radio
propagation physics, or models that resort to a predefined
connectivity scenario.

The {\em Edge Model} uses the {\em Communication Model} for providing a
graph representation of the network by giving access to the direct
neighbors of a node. This has two major implications. First, it allows
for simple centralized algorithms that need information on the
communication graph. In this, Shawn differs from Ns-2 and other
simulators, for which the check for connectivity must be based on sending
test messages. The second point is the exchangeability of edge models:
Simulations of relatively small networks may allow storing the
complete neighborhood of each node in memory and will thus provide
extremely fast answers to queries. However, huge networks will impose
impractical demands for memory; therefore, an alternative edge model
trades memory for runtime by recalculating the neighborhood on each
request, or only caches a certain number of neighborhoods.

While the {\em Communication Model} decides whether two nodes can
communicate as a matter of principle, the {\em Transmission Model}
determines the properties of an individual message transmission. It
can arbitrarily delay, drop or alter messages. This means that 
when the runtime of
algorithms is not in question, a simple transmission model without
delays is sufficient. A more sophisticated model may account for contention,
transmission time and errors.

Different implementations of these models can significantly alter the
behavior of the simulation. This can either mean changing the
behavior of the virtual world or modifying the requirements of the
simulation. An example of a change to the virtual world is the use of
a different {\em Transmission Model}, e.g., using random message
dropping. Depending on the size of the simulated world, a change in
the implementations of e.g. the {\em Edge Model} may substantially
alter the performance and the requirements of the simulation.

More specialized models provide data for simulations. Currently Shawn
ships with {\em Random Variable} and {\em Node Distance Estimate}
models. Random variables are needed very often in simulations for modeling
the real-world behavior. With the introduction of random
variable as models, algorithms can be tested with different underlying
random variables without the need of being aware of the change. {\em Node
  Distance Estimate} implementations are used to mimic distance
measurements for, say, localization algorithms.

\subsection{Sequencer}
\label{sec:seq}
The sequencer is the central coordinating unit in Shawn. It configures
the simulation, executes tasks sequentially and drives the simulation.
It consists of the Simulation Controller, the Event Scheduler and the
straightforward, yet powerful, concept of Simulation Tasks.

The purpose of the Simulation Controller is to act as the central
repository for all available model implementations and to drive the
simulation by transforming the configuration input into parameterized
calls of Simulation Tasks. These are arbitrary pieces of code
that can be configured and run from the simulation's setup files.
Because they have full access to the whole simulation, they are able to
perform a wide range of jobs. Example uses are the steering of
simulations, gathering data from individual nodes or running
centralized algorithms.  Finally, the Event Scheduler triggers the
execution of events that can be scheduled for arbitrary discrete
points in time.

\subsection{Simulation environment}
\label{sec:env}
The simulation environment is the home for the virtual world in which
the simulation objects reside. All nodes of a simulation run are
contained in a single world instance. The nodes themselves serve as a
container for so-called Processors, which are the real work horses of
the simulations; they process incoming messages, run algorithms and
emit messages.

Shawn features persistence and decoupling of the simulation
environment by introducing the concept of {\em Tags}. They attach both
persistent and volatile data to individual nodes and the world. They
decouple state variables from member variables, thus allowing for an
easy implementation of persistence. Another benefit is that parts of a
potentially complicated protocol can be replaced without modifying
code, because the internal state is stored in tags and not in a special
node implementation.

\section{Case Study}
\label{sec:casestudy}
To demonstrate Shawn's performance gain, we now present a comparison to
Ns-2. We ran a number of simulations of a subroutine that is used in
certain time synchronization protocols.  Here every node periodically
broadcasts a message containing time stamps that is converted at the
receiving node. Meanwhile, overhead induced by the time stamps is
measured. A total of 380 messages is sent by each node. This provides
insight on the simulator's ability to dispatch a large amount of
traffic.

The result of this comparison is not surprising, because Ns-2 does a
lot more detailed computations than Shawn to arrive at the same
results. This shows that Ns-2 and others cannot compete with Shawn in
its domain.

Table~\ref{tab:results} shows the runtime and memory consumption of
Ns-2 and Shawn in different setups. The environment consists of a
square area whose size is the specified multiple of the nodes'
communication range. The node density describes the average number of
nodes within a broadcast area.

The first thing to notice is that Ns-2 hits the one-day barrier for instances
that Shawn finishes in less than one minute with considerably
smaller memory footprint. The fifth line refers to a simulation run
using the ``Simple'' edge model in which neighborhoods are not cached at
all and hence the simulation uses more time and less memory. In all
other runs, the ``List'' edge model is used, which completely caches
all neighborhoods. This is one example for which the choice of model can
trade memory versus runtime. The last three lines show networks of
huge size, respectively, huge neighborhoods, that only Shawn can handle
in reasonable time.

\begin{table*}
\noindent
\begin{tabular}{|r|r|r|rr|rrl|} 
\hline 
  
& 
& 
& \multicolumn{ 2}{|c}{{\bf Ns-2}}
& \multicolumn{ 3}{|c|}{{\bf Shawn}} \\ 
  \multicolumn{1}{|p{1.5cm}}{Number of Nodes} 
& \multicolumn{1}{|p{1.5cm}}{Environment Size}
& \multicolumn{1}{|p{1.5cm}}{Node Density}    
& \multicolumn{1}{|p{1.5cm}}{CPU Time (H:M:S)}
& \multicolumn{1}{p{1.5cm}|}{Memory Usage (MBytes) }
& \multicolumn{1}{p{1.5cm}}{CPU Time (H:M:S) }
& \multicolumn{1}{p{1.5cm}}{Memory Usage (MBytes)}
& \multicolumn{1}{p{1.5cm}|}{Edge Model} \\ 
\hline 
100 & 10x10 & 3.1 & 00:00:15 & 14.8 & 00:00:01 & 1.9 & List \\ 
\hline 
500 & 10x10 & 15.7 & 00:22:59 & 53.9 & 00:00:01 & 2.9 & List \\ 
\hline 
1,000 & 10x10 & 31.4 & 01:59:36 & 106.0 & 00:00:04 & 4.5 & List \\ 
\hline 
2,000 & 10x10 & 62.8 & 25:36:13 & 224.0 & 00:00:19 & 8.6 & List \\ 
\hline 
25,000 & 10x10 & 785.4 & & & 19:45:48 & 122.9 & Simple \\ 
\hline 
30,000 & 10x10 & 942.5 & & & 01:34:47 & 757.6 & List \\ 
\hline 
200,000 & 80x80 & 78.5 & & & 03:27:49 & 891.0 & List \\ 
\hline 
300,000 & 173.2x173.2 & 31.4 & & & 04:47:46 & 855.5 & List \\ 
\hline 
\end{tabular} 
\caption{Comparison of running time and memory usage between Shawn and Ns-2.}
\label{tab:results}
\end{table*}

\section{Conclusions}
\label{sec:conclusion}
We have presented Shawn, an open-source discrete event simulator for
sensor networks with huge numbers of nodes.  By reviewing existing
simulators, we have identified a previously uncovered gap in
simulation domains. By means of a simple case study we have
demonstrated what Shawn's strengths are and how it fills the described
gap. We have described the differences between Shawn and its
competitors, its unique features and how users can benefit from its
application.

\section{Future work}
\label{sec:future}
A crucial point in the future will be to provide more model
implementations. Our current plans are to supply different mobility
models and fine-grained communication and transmission models. We
strongly encourage the open-source community to participate in this
process and to enhance Shawn by contributing to its growth.

Another planned improvement is a better interface for discrete
combinatorics. Existing libraries such as CGAL~\cite{CGAL} and BOOST
Graph~\cite{BOOST} provide sophisticated data structures and
algorithms for computational geometry and graph theory. By making
Shawn's internal network structure visible to these libraries we can
immediately leverage their code base. Furthermore, we want to support
the data formats of Ns-2 in order to be able to process existing
scenarios.

\section{Acknowledgements}
\label{sec:ack}
This work is part of the SwarmNet project
(\url{http://www.swarmnet.de}) and funded by the German Research
Foundation (DFG) under grants Fe 407/9-1 and Fi 605/8-1.  We thank
Tobias Baumgartner and Andr{\'e} Steinert for their commitment in the
preparation of this paper.

\bibliographystyle{IEEEtran}
\bibliography{references}

\end{document}